******************************************************************

\documentstyle[12pt]{article}
\newcounter{mycount}

\newcommand{\be}{\begin{eqnarray}}
\newcommand{\ee}{\end{eqnarray}}
\newcommand{\bfl}{\begin{flushleft}}
\newcommand{\efl}{\end{flushleft}}
\newcommand\ie {{\it i.e. }}

\newcommand\half{\frac 1 2 }
\newcommand\noi{\noindent}

\begin{document}

\centerline{\Large\bf QED WITH UNEQUAL CHARGES}
\vskip 2mm
\centerline{\large\bf  a study of spontaneous $Z_n$ symmetry breaking  }
\vspace* {-35 mm}
\begin{flushright} USITP-94-09 \\
SUNY-NTG-94-23 \\
May 1994
\end{flushright}
\vskip 0.9in
\centerline{  \bf T.H. Hansson$^a$, H. B. Nielsen$^b$ and
  I. Zahed$^c$  }
\vskip 25mm\noi
\centerline{\bf ABSTRACT}
\vskip 5mm

We study two-dimensional QED with unequal charges at finite
temperature, and show that there is a phase with a
spontaneously broken $Z_n$ symmetry. In spite of this, we were not able to
establish the presence of domain walls. The relevance for QCD in higher
 dimensions is discussed.

\vfil\noi
$^a$
Institute of Theoretical Physics, University of Stockholm \\
Vanadisv\"agen 9, S-113 46 Stockholm, Sweden \\
hansson@vana.physto.se
\vskip 2mm\noi
$^b$ Niels Bohr Institute, \\
Blegdamsvej 17, Copenhagen $\rlap/{\rm O}$, Denmark \\
hbech@nbivax.nbi.dk
\vskip 2mm \noi
$^c$Nuclear theory group, Department of physics, SUNY at Stony Brook \\
Stony Brook, New York, 11794, USA \\
zahed@sbnuc1.phy.sunysb.edu

\vskip 3mm \noindent
${^a}$Supported by the Swedish Natural Science Research Council. \\
${^c}$Supported in part by the Department of Energy under Grant \\
 No. DE-FG02-88ER40388.

\eject

\newcommand\nc {$N_c$}
\newcommand\zn {$Z_n\ $}
\newcommand\lan {\langle}
\newcommand\ran {\rangle}
\newcommand\pad {{\cal D}}
\newcommand\rarr {\rightarrow}

\noi {\bf 1. Introduction}
\vskip .5cm

Recently there has been some controversy concerning the \zn phase
structure of QCD, with and without dynamical quarks. The notion that \zn
symmetry is spontaneously broken in the high temperature phase of
pure Yang-Mills theory \cite{POL,SUS,WEI}, is
strongly supported by lattice Monte Carlo simulations \cite{LAR,KUT,SVE}.
It is thus natural to believe that in a large enough system, above the
transition temperature, different \zn phases can coexist and be separated by
domain walls, much like the domains in a ferromagnet. The surface
tension of such domain walls has been calculated using effective action
methods,\cite{BHA} and lattice simulations\cite{DIX,KUP}.
This standard scenario has recently been challenged. It was noticed
in \cite{BEL,CHE}, that if dynamical fermions were included,
the thermodynamics in the hypothetical \zn
bubbles is marred with contradictions such as negative entropy.
This has led some authors to question the whole concept of spontaneously broken
\zn symmetry \cite{BEL,CHE}, even going so far as to argue that what has been
observed in the Monte Carlo simulations are in fact lattice artifacts
\cite{SMI}. The arguments in this last paper are largely based on analogies
with
two-dimensional models and four-dimensional QED. In this paper we shall also
consider two dimensional models but with an essential new ingredient: we will
allow for particles with different charges.
As we shall show, this gives rise to a \zn
symmetry, very much like the center symmetry in pure Yang-Mills theory, but
with the possibility of making well controlled calculations. In such a
model (essentially the Schwinger model) we can establish
spontaneous symmetry breaking at all temperatures nonperturbatively.
We shall see that
this result appears to follow from some very general properties of a
screening system, and
we will argue that it is expected also in higher dimensional Yang-Mills
theories above $T_c$.
Concerning domain walls, we shall give strong arguments for why they
cannot exist in the two-dimensional model and again argue that the same may
hold
true in higher dimensions and for non-Abelian fields.

\vskip 1cm\noi
{\bf 2. \zn symmetry in models with unequal charges }
\vskip .5cm
\noi
Consider the following Lagrangian
\be
{\cal L} = -\frac 1 4 F^2- \overline\psi_h[i\gamma^\mu(\partial_\mu -
ieA_\mu) - M]\psi_h - \overline\psi[i\gamma^\mu(\partial_\mu -
iNeA_\mu) - m]\psi
\ee
where we shall take $M\rarr \infty$ and consider the heavy particles (described
by
$\psi_h$)  only as
probes of the dynamical theory defined by the light fermions and the gauge
fields.
We shall use the imaginary time approach to finite temperature and thus
consider the Euclidian field theory defined on $x^0 \in [0,\beta]$ with proper
periodicity conditions. The relevant Polyakov loop operators corresponding to
the light and heavy fields are given by ,
\be
W(x) &=& e^{ie\int_0^\beta dx^0\, A^0(x^\mu)} \\
W_l(x) &=& e^{ieN\int_0^\beta dx^0\, A^0(x^\mu)} \nonumber
\ee
respectively. It now follows that constant $A^0$ fields in the interval
$[0,2\pi/\beta [$ are gauge nonequivalent since they correspond to different
values for $W(x)$. All other constant $A^0$ are related to points in this
interval via a proper gauge transformation of the type $A^0 \rarr A^0 + 2\pi
n/\beta$. Now we note that the dynamical sector of the theory, \ie the light
fermions and the gauge fields, have an additional \zn symmetry corresponding
to
\be
A^0 \rarr A^0 + \frac {2\pi} \beta \frac k N \ \ \ \ \ k=1, 2, ....N-1
 \ \ \ \ \ . \label{03}
\ee
This \zn symmetry is of the same nature as the \zn symmetry in hot Yang-Mills
theories, if we think of the light fermions as the gluons, and the heavy ones
as the quarks.

To decide whether this \zn symmetry is manifest, or spontaneously broken, we
shall use the standard method employed to study magnets: we add a small
symmetry breaking term, $H_{sb}$ to the Hamiltonian (corresponding to a
magnetic
field in a spin system) and ask whether an ordered state is left after $H_{sb}$
is taken to zero. The natural term in our case is

\be
H_{sb} = \int d^dx\, \epsilon(x) W(x)  \ \ \ \ \ ,
\label{break}
\ee
where the small parameter $\epsilon$ will be specified later.
To establish spontaneous symmetry breaking we use a constant
$\epsilon$, and later we make it $x$-dependent in order to study domain
walls.

\vskip 1cm\noi
{\bf 3. The Schwinger model}
\vskip .5cm

We shall now consider the two-dimensional case, \ie the Schwinger model. To
regulate the infrared divergences we compactify space to a circle and thus
define the Euclidian field theory on a torus $L\times\beta$. Following the
treatment in \cite{SAC} we decompose the gauge field as
\be
eA^0 &=& \frac {2\pi} \beta h_0 - \partial_1 \phi + \partial_0 \lambda \\
eA^1 &=& \frac {2\pi} \beta h_1 + \partial_0 \phi + \partial_1 \lambda
         \nonumber \ \ \ \ \  ,
\ee
where both the Schwinger field $\phi$ and the pure gauge field $\lambda$ are
strictly periodic on the torus.
Clearly $h_\mu \rarr h_\mu + 2\pi n_\mu$ ($n_\mu$ integers) are proper gauge
transformations so we can make the restriction $h_\mu \in [0,1 [$ in the
integration over the gauge fields.  The \zn symmetry (\ref{03}) of the
dynamical
fields now correspond to the transformation
\be
h^0 \rarr h^0 + \frac {2\pi} \beta \frac k N \ \ \ \ \ k=1, 2, ....N-1
\ee
on the constant part of the timelike gauge potential.

By analogy with the calculation of the magnetization in a spin system,
we shall compute the expectation value of a Polyakov loop with
(heavy) unit charge particles via the following limiting procedure,
\be
\lan W(x)\ran_\epsilon = \lim_{\epsilon\rarr 0} \lim_{L\rarr \infty}
 \frac {\int  \pad A_\mu\pad\psi\pad\overline\psi e^{-S_{S}-S_{sb}} W(x) }
  {\int  \pad A_\mu\pad\psi\pad\overline\psi e^{-S_{S}-S_{sb}} }
= \lim_{\epsilon\rarr 0} \lim_{L\rarr \infty} \frac {\cal N}{\cal D}
\label{pol}
\ee
Following \cite{SAC} we can, for the case $m=0$, integrate out the
fermions from the Schwinger model action $S_S$ and change variables from
$A_\mu$ to $h_\mu$ and $\phi$. Expanding the symmetry breaking term we get
 \be
{\cal N} = \int d^2h\, |\Theta(h_\mu)|^2
   \sum_{n=0}^\infty \frac {\beta^n} {n\!} \langle W(x)\Pi_{i=1}^n
\int dx_i \epsilon (x_i) W(x_i ) \rangle ,
\label{07}
\ee
and a similar expression for $\cal D$. In (\ref{07}), $\Theta(h_\mu)$ is an
expression involving the Jacobi theta
function, $m_S^2 = (Ne)^2/\pi$ the Schwinger mass, and the expectation value
is over the $\phi$-field with weight
\be
e^{-\Gamma[\phi]} = e^{-\half \int d^2x\, \phi \Box(\Box -  m_S^2 )\phi(x)}
\label{wei}
\ee
Substituting
 \be
W(x) = e^{i[2\pi h_0 - \partial_1\int_0^\beta dx^0\, \phi(x)] } \ \ \ \ \ ,
\label{ppol}
\ee
in (\ref{07}), we can do the $h$-integration. For the $n$-th term and
large $L$ we get
 \be
\lim_{L\rarr \infty} \int d^2h\, |\Theta(h_\mu)|^2 e^{2\pi i (n+1) h_0} =
\frac 1 {\sqrt{2N\tau}} \sum_{k=o}^{N-1} e^{-2\pi i \frac {k(n+1)} N }
\ee
where $\tau = L/\beta$.
Note that the $k$-sum in this expression is nothing but a periodic Kronecker
delta modulo $N$, so the net effect of the $h_0$ integration is to select the
terms in (\ref{07}) where the number of loops is a multiple of $N$. The
physical
interpretation is obvious - in these cases the total charge is $lNe$ which can
be compensated by $l$ dynamical fermions of charge $Ne$ from the heat bath.
Other values of the charge would violate Gauss' law and are forbidden. It is
important to notice that it is the $h_0$ integration that implements Gauss'
law globally (while the $\pad \phi$ integration implements it locally).

The $\pad\phi$ integration is again Gaussian and gives the following result for
the correlation functions in (\ref{07}) \cite{SAC}
\be
\lan W(x_1)W(x_2)....W(x_n) \ran_\phi = (\lan W(x) \ran_\phi)^n
\prod_{i<j} \frac {\lan W(x_1)W(x_2) \ran_\phi}{(\lan W(x) \ran_\phi)^2}
\ee
where
\be
\lim_{L\rarr \infty} \lan W(x) \ran_\phi &=&
       e^{-\beta F_1} = e^{-\beta \frac \pi {4N^2} m_S} \\
\lim_{L\rarr \infty}  \lan W(x_1)W(x_2) \ran_\phi  &=& (\lan W(x) \ran_\phi)^2
\exp\left( -\frac {\beta\pi m_S} {2N^2} e^{-m_S|x_1^1 - x_2^1|} \right)
\ee
For pair separations $|x_i^1 - x_j^1|>1/m_S$ the correlations between the
Polyakov lines can be neglected so that

\be
\lan W(x_1)W(x_2)....W(x_n) \ran_\phi \sim (\lan W(x) \ran_\phi)^n
 \ \ \ \ \ .
\ee
This allows us to reexponentiate the sums in (\ref{07}) to get (up to
exponentially small corrections),
\be
\lan W(x)\ran_\epsilon = \lim_{\epsilon\rarr 0} \lim_{L\rarr \infty}
e^{-\beta F_1} \frac
 { \sum_{k=0}^{N-1}  e^{-2\pi i \frac k N}
\exp\left(\beta e^{-\beta F_1 } \int dx^1\epsilon(x^1)e^{-2\pi i \frac k N}
 \right) }
{\sum_{k=0}^{N-1}
\exp\left(\beta e^{-\beta F_1} \int dx^1\epsilon(x^1)
e^{-2\pi i \frac k N} \right) }
\label{15}
\ee

First, consider the case where $\epsilon(x)$ is space independent and
proportional to a \zn element, that is
 \be
\epsilon(x) = \lambda e^{+2\pi i \frac m N} \ \ \ \ \ ,
\label{16}
\ee
with $\lambda$ small and real. Inserting  (\ref{16}) into (\ref{15})
yields
 \be
\lan W(x)\ran_\epsilon = e^{-2\pi i \frac m N} e^{-\beta F_1} \ \ \ \ \ .
\ee
Thus we have established that the \zn symmetry is spontaneously broken.
The above vacuum average for the Polyakov line is consistent
with the usual identification with the free energy of a heavy fermion. Indeed,
 \be
-\frac 1{\beta}{\rm log} \lan W(x)\ran_\epsilon = F_1 + i \frac {2\pi}{\beta}
\frac {m}N
\label{free}
\ee
$F_1= (\pi m_S)/(4N^2)$ is  the self-energy of a heavy and screened
fermion,
\be
F_1 = \frac {e^2}{2}\int \frac {dq}{2\pi}\frac 1{q^2+m_S^2}
= \frac {\pi m_S} {4N^2}
\ee
The imaginary part is a relique of the constant (large) gauge transformation
along the temporal direction. The energy of a charge is only defined
modulo a constant shift in the potential. The imaginary character of the shift
is related to the fact  that the calculation was performed in Euclidean
space, where the potential is imaginary. Note that at zero temperature,
the large Wilson loops display a perimeter law, with coefficients $\kappa_l =
N^2\kappa_h = N^2 F_1$, for light and heavy probes respectively.
This follows since $\kappa_h$ and $\kappa_l$ are just the Coulomb
self-energies. The perimeter fall off is slower for the
heavy probes since the screening by the light charges is weaker.
This result is generic and carry to four dimensions.
In the latter, however, a pertinent renormalization is needed.

The symmetry breaking occurs at any temperature which is somewhat
surprising, since at
very high temperatures we are essentially dealing with a one-dimensional
statistical
system, so there should be no symmetry breaking.
The resolution of this puzzle lies in
the non-local nature of the gauge interaction. Technically, this comes
about via the
integration over the variable $h_0$ in (\ref{07}). Since $eh_0/\beta$
is the flux through the cylinder with radius $\beta$ on which the fields are
defined,  $h_0$ is directly related to the value of the Wilson loop.
The measure depends non-trivially
on $h_0$ via $\Theta(h_\mu)$, and the net result of the integration, is to
implement Gauss law, and to correlate the Polyakov loops  over long distances.

What about domain walls? In analogy with classical statistical
mechanics, we would expect domain wall formation at low temperatures, where we
effectively have a two dimensional system with a broken discrete symmetry. At
high temperature, we would not expect  domain walls. However,
even in a one dimensional
spin system, we expect both symmetry breaking and domain wall formation
if we never
totally take the external fields to zero, but keep them small and finite.
To study domain wall formation, we subdivide the circle into two regions
of length
$L_1=x_1L$ and $L_2=x_2L$ ($x_1+x_2 = 1$) with {\em constant}
$\epsilon_1 = \lambda_1 z_1$ and
$\epsilon_2= \lambda_2 z_2$ respectively, where $z_1$ and $z_2$ are two
different elements of the center \zn. A calculation similar to the above
gives,
\be
\lan W(x)\ran_\epsilon = e^{-\beta F_1}z^*_k
\ee
where $z_k$ is the center element which maximizes
$\rm{Re}(z_k^*z_1\lambda_1 x_1+ z_k^*z_2 \lambda_2 x_2)$. For the case of $N=2$
we get the simple result
\be
\lan W(x)\ran_\epsilon = e^{-\beta F_1}
\left[ z_1^* \theta(\lambda_1 x_1 - \lambda_2 x_2 ) +
       z_2^* \theta(\lambda_2 x_2 - \lambda_1 x_1 )\right] \ \ \ \ \ ,
\label{20}
\ee
where  $\theta$ is a step function.
In this case the first term survives for $L_1 > L_2$, while for $L_2 >L_1$
it is the second one, so the largest domain  dominates the
thermodynamical limit\footnote{We note that for finite
volumes domain walls can be triggered, but then the whole concept of
spontaneous symmetry breaking is lost. This point might be relevant for the
lattice simulations.}.
This is also true for arbitrary $N$ as long as one domain is
sufficiently much larger than the other. This argument can
easily be generalized to the case of several domains.
We conclude, that although the thermal state
spontaneously breaks \zn symmetry, domain walls cannot be $forced$ even
by nonvanishing ''magnetizations''. It is natural to believe that the symmetry
breaking merely expresses  that the high temperature phase screens.
The \zn shift in the order parameter reflects  the gauge dependence
of the screening mass of a single charge under constant temporal gauge
transformations, which  is not fixed in the functional integral.
This is consistent with the non-existence of domain walls, since their presence
would convey a physical meaning to the phase of the partition function.

\vskip 1cm \noi
{\bf 4. The massive Schwinger model}
 \vskip .5cm

So far, we have assumed that the light charges were massless. As a result,
the topologically nontrivial sectors, corresponding
to nonvanishing electric flux through the surface of the torus, did not
contribute to the functional integral due to the presence of zero modes
(which makes the fermion determinant vanish). It is an interesting property of
the Schwinger model, that the presence of a fermion mass drastically alters the
infrared behaviour. This was first observed by  Coleman, Jackiw and Susskind
\cite{COL},
who showed, using bosonization techniques, that the mass induces an
effective string
tension proportional to the fermion condensate. This string tension is not
to be mistaken for the trivial linear Coulomb force which is present in
classical 1+1 dimensional electrodynamics.
Below we indicate how this result can be rederived
using functional integral methods. Here the pure
quantum mechanical origin of the
confining force is manifest since the whole effect
comes from integrating over the
topologically non-trivial sectors of the gauge field.

To probe the character of the two dimensional vacuum in the presence of a
small mass term, consider the correlation function between a
Polyakov-anti-Polyakov line. To leading order in $m$ we have,
\be
\lan W(x) W^+(0) \ran_m \approx \lan W(x) W^+(0) \ran_0 +
m \int d^2y\,\lan W(x) W^+(0) i\psi^+\psi (y)\ran_0^c \\
\approx \lan W(x) W^+(0) \ran_0
\exp\left[-m \int d^2y\, \frac { \lan W(x) W^+(0) i\psi^+\psi (y)\ran_0^c }
{ \lan W(x) W^+(0) \ran_0^c } \right] \nonumber
\label{1}
\ee
where the Green's function in the exponent is connected. Using arguments
similar
 to
the ones given by Sachs and Wipf \cite{SAC},
we obtain for the (unsubtracted) correlation function
\be
&&\int d^2y \lan W(x) W^+(0) i\psi^+\psi (y)\ran_0^c   \\
&&\qquad = \half \lan W(x) W^+(0) \ran_0 \,\,\lan i\psi^+\psi\ran \cos
\left(R(x,y)-\frac {2\pi} L x^1 - \vartheta \right) \nonumber
\label{2}
\ee
with
\be
R(x,y)=\frac 1 {NV}\int_0^\beta dx^0\, \left[K'(x^0,x^1;y)-K'(x^0,0;y)\right]
\\
\nonumber
K'(x;y) = \frac \partial {\partial x^1} \langle x| \frac 1 {\Box^2 -
     \Box m_S^2 } |y\rangle \sim \frac 1 {m_S^2}
 \langle x| \frac 1 \Box |y\rangle
\label{3}
\ee
and where $\sim$ denotes the leading contribution as $L\rightarrow\infty$.
We also introduced the vacuum angle $\vartheta$ which is the variable conjugate
to
the total flux through the surface of the torus.
By going to momentum space, it is easy to verify that,
\be
\int_0^\beta dx^0\,  \langle x| \frac 1 {\Box} |y\rangle = -\frac L {2\pi^2}
\sum_{n\neq 0}\frac {\cos{\frac {2\pi} n (x^1-y^1)}} {n^2}
\sim \half |x^1-y^1| \ \ \ \ \ .
\label{33}
\ee
Substituting (\ref{33}) into (25) gives
\be
R(x,y) \sim \frac {2\pi} n \theta(x^1-y^1)
\ee
where $\theta(x)$ is a step function.
Inserting this result into (\ref{1}) yields
\be
\lan W(x) W^+(0) \ran_m \sim \lan W(x) W^+(0) \ran_0 \,\, e^{-\sigma \beta x}
\label{6}
\ee
with a "string tension"
\be
\sigma = m \left( i \lan \psi^+ \psi\ran_0\right) \left(
{\rm cos} (\vartheta - \frac {2\pi}N ) -{\rm cos} (\vartheta ) \right)
\label{7}
\ee
Since our calculation is only to lowest order in $m$, we cannot from this
result conclude that the presence of a small mass really induces a string
tension. We note, however, that the above lowest order result for $\sigma$
is exactly that obtained  by
Coleman, Jackiw and Susskind \cite{COL}. Although we have not bothered, we
believe that it is straightforward to
check that the  subleading contributions in $m$
to $\lan W(x) W^+(0) \ran_m$,
will not give any corrections to the string tension. (This is also more or
less obvious for dimensional reasons.
Notice that the Minkowski space condensate
$\lan \overline\psi \psi \ran = i\lan \psi^+ \psi\ran$, which
means that the fermion condensate is purely imaginary in Euclidean space
The calculations performed in \cite{SAC} shows that it is real
in Euclidean space. The missing $i$ comes from the arbitrariness in the
relative phase between the zero modes and their
conjugates in Euclidean space.)
As expected, integer charges $N=1$ do not confine in the $\vartheta =0$
(modulo $\pi$) sector, but also note that the string tension
vanishes for $\vartheta =\pi/N$ (modulo $2\pi$).

What happens to the spontaneous breaking of \zn in this case? Since this
phase confines, we  expect the expectation value of the Polyakov line
to vanish even in the presence of a small symmetry breaking term.\footnote{
It is amusing to note that the role of the mass
is analogous to the role of temperature in QCD. As we dial down the mass, we
move from a confining to a screened phase. The transition point occurs at
$m=0$.}
To see that this is indeed the case, we note that in the presence of a mass
term, the expectation value of the symmetry breaking term can be obtained
from the following partition function,
\be
{\cal Z} [\epsilon] =
\lan e^{\int dx \epsilon (x) W (x)}\ran_{m,\vartheta} \sim &&
\sum_{n=0}^{\infty} \frac 1{n!}\lan \prod_{i=1}^n \int dx_i \epsilon (x_i)
W(x_i)\ran_{0,0} \nonumber \\&&
e^{ - \sum_{j=2}^{n} |x_j-x_{j-1}| \sigma_{j-1}}
\left( \frac 1N \sum_{k=0}^{N-1} e^{-i\frac {2\pi}N nk}\right)
\label{8}
\ee
where the space dependent string tensions are given by
 \be
\sigma_{j} = m \lan i\psi^+\psi\ran
|{\rm cos}(\vartheta - j\frac{2\pi}N )-{\rm cos}(\vartheta )|
\ \ \ \ \ .   \label{9}
\ee
For $m=0$, ${\cal Z}$ is nothing but ${\cal D}$ in (\ref{pol}).
The last term in (\ref{8}) shows that only the string of terms for which
$K= Np$ where $p$ is an arbitrary integer contribute to the sum. Let
$Np$ be such a string. Its contribution to the partition function ${\cal
Z}[\epsilon ]$ is
\be
{\cal Z}_p [\epsilon]
\frac{(Np)!}{(N!)^p p!}
\left[e^{-N\beta F_1}
\prod_{i=1}^N \int dx_i\, \epsilon (x_i)
{\rm exp}\left( - \sum_{j=2}^{K} |x_j-x_{j-1}| \sigma_{j-1}\right) \right]^p
\label{10}
\ee
The combinatorics follows from the number of possibilities to split $Np$
terms into $p$ bins, with $N$ identical terms in each bin. If we assume
that $\epsilon (x) =\lambda z$ where $z$ is a constant element of \zn,
and evaluate the integrals within each ''cluster'' in terms of an an
average string tension $\sigma$, then
\be
\frac 1 {N!} \prod_{i=1}^N \int dx_i
{\rm exp}\left( - \sum_{j=2}^{K} |x_j-x_{j-1}| \sigma_{j-1}\right)
= \frac L{\sigma^{N-1}}
\label{11}
\ee
Trading the summation over $K$ in (29) for a summation over $p$ the
partition function (\ref{10}) becomes
 \be
{\cal Z} [\epsilon] = \sum_{p=0}^\infty {\cal Z}_p [\epsilon]
 = {\rm exp}\left( \frac{L\epsilon^Ne^{-N\beta
F_1}}{\sigma^{N-1} }\right) \ \ \ \ \ .
\label{12}
\ee
For $N\geq 2$ the average value of the Polyakov line is given by
 \be
\lan W(x)\ran_{m,\vartheta} ={\lim_{\lambda\to0}\lim_{L\to\infty}}\frac 1{Lz}
\frac{\partial}{\partial\lambda}{\rm log}{\cal Z} =
\frac{e^{-N\beta F_1}}{(N-1)!}\,\,z^*\,\,
\lim_{\lambda\to0}\left(\frac {\lambda}{\sigma}\right)^{N-1} = 0
\label{13}
\ee
which is the expected result. For $N=1$ the loop can be screened by a
particle from the heat bath and $W$ acquires an expectation value.

Note that when $\vartheta =\pi/N$, the string tension between
the Polyakov-anti-Polyakov line cancels. This phase, which
is not totally screening as $(N-1)$ strings of Polyakov lines are
still confined, is discussed in the Appendix.

\vskip 1cm\noi
{\bf 5. Summary and Conclusions}
\vskip .5cm

We have considered an extended version of the Schwinger model with both
non-dynamical (heavy) charge $e$, and dynamical (light) charge
$Ne$, fermions. There is a \zn symmetry of the dynamical sector of the
theory due  to  constant $U(1)$ gauge configurations that change the
Polyakov loops for the heavy fermions by a \zn element while leaving those
of the light fermions invariant. This situation is reminiscent of QCD if we
think of the light fermions as gluons and the heavy fermions as quarks.

For massless
dynamical fermions, the system is screening at all temperatures,
and the \zn symmetry is  spontaneously broken. In spite of this,
no domain walls can be formed in the thermodynamical limit. For massive
fermions an effective string tension $\sim m$ is generated by quantum
effects, just as in 3 or 4-dimensional QCD. Note that this is not the case
in the t'Hooft model where confinement is essentially classical, and the
string tension survives since particle production is down by $1/N_c$. It is
also
interesting to note that the field configurations that give rise to the
string tension in the massive Schwinger model all carry non-zero
topological charge. In the $m\neq 0$ case, the theory is confining, and, as
expected, there is no spontaneous breaking of the \zn symmetry.

What are the consequences for Yang-Mills theory in higher dimensions?
The essential ingredient in our calculation was the integration over the
constant piece  $h_0$ in the  time-like potential. In the absence of a
symmetry breaking term in the action, this integration gives zero for the
expectation value of the Polyakov loop. This is an exact result independent
of the size of the system and in fact a consequence of Gauss law.
We believe that this will carry over
without change to higher dimensions and non-abelian symmetries, since the
constant modes are present also in that case. Also, since they take values in
the group center, the non-abelian nature should play no role. We thus
claim, that without the presence of a symmetry breaking term the value of
the Wilson loop is always identically zero. This is clearly consistent with
the general lore about spontaneous symmetry breaking.

How does this compare with the lattice Monte Carlo simulations
\cite{LAR,KUT,SVE}? In particular, what is it in these simulations that play
the role of a symmetry breaking term? In fact, there is nothing.
As is commonly assumed, we believe that the observed jumps
in the phase of the Polyakov loop is a finite volume effect
that will disappear in the thermodynamical limit.
We also believe that the hysterisis observed in finite volumes is
caused by the way the simulations are actually performed.
Starting from some
initial configuration, the system is equilibrated via a local updating
procedure, that ''freezes'' the system at a particular \zn value, except for
few configurations for which a jump occur. We can imagine
changing the updating procedure by a rule that also average over all the
gauge-equivalent \zn configurations. Such an updating procedure, that would
be as good as the normal one for any zero $N-$ality quantities (like
$\langle W W^\dagger\rangle$ or operators containing only gluons), and would
enforce Gauss law globally, would of course give zero for the Polyakov loop.
In other words, the difference between our
Schwinger model calculation and the present lattice simulations is that
the former explicitly averages over all gauge copies, while the latters
(in the thermodynamical limit) do not. Of course, with such a new updating
procedure an explicit breaking term such as (4) could be added
and then removed by the condition $\epsilon V_4 << 1$. Such a procedure
would totally paralell our continuum calculations, and we think that in this
case our conclusions hold.

Will our results about domain walls also carry over to real QCD ? Again we
think they will. From the above it is clear that the \zn symmetry breaking
observed in the lattice simulations, and in our calculation, is in a sense
spurious since it depends on how the calculation is performed.
The apparent breaking of \zn symmetry lies in the non-local nature of
the gauge interaction as stressed above. This does
not mean that the expectation value of the Polaykov loop is not an
interesting variable, only that its phase does not carry any physical
meaning (while the real part is related to the free energy).  If the
phase of $\langle W \rangle$ does not carry physical meaning
there is no reason to believe that there can be domain walls, and indeed
our calculation demonstrates that there are none. Since we cannot carry out
the calculations in real QCD, we can clearly not prove this point, and
the presence of domain walls have been claimed in the literature, using
lattice simulations\cite{DIX,KUP} and analytical calculations\cite{BHA}.
In both these calculations a domain wall is forced by picking appropriate
boundary conditions (either by introducing a \zn flux through a plaquette on a
toroidal lattice, or by simply picking different asymptotic values
for $h_0$). Although such boundary conditions certainly can be introduced
in a calculation, it is not clear that they would be relevant for a real
physical system. For instance, in the Schwinger model on the torus, no such
domain wall can occur if we require the field configurations to be
non-singular (on the lattice this can be avoided by piercing a plaquette
with a \zn flux). Clearly it would be very pleasing if our Schwinger model
results carried over to QCD, since that would solve the apparent
contradictions related to \zn bubbles in the presence of
fermions\cite{BEL,CHE}. We note that our conclusions are quite similar
to the ones reached by Smilga \cite{SMI}.

\vglue 1.5cm
{\bf \noindent  Acknowledgements }
\vskip .5cm
\noi
We would like to thank Larry McLerran and Andrei Smilga
for an interesting discussion.
This work was partially supported by the US DOE grant DE-FG-88ER40388.

\newpage\noi
{\large\bf Appendix : The case $\vartheta =\pi/N$ }
\renewcommand{\theequation}{A.\arabic{equation}}
\setcounter{equation}{0}
\vskip .5cm
As pointed out in section 4, the partition function (30) can be thought of
as an admixture of strings of length $(N-1)$ with  fugacity
\be
\xi_1 = \frac{L\epsilon^{N-1} e^{-(N-1)\beta F_1}}{(N-1)!\sigma^{N-2}}
\label{14}
\ee
and free charges of fugacity
 \be
\xi_2 = L\epsilon e^{-\beta F_1}
\label{155}
\ee
The partition function for this system can be readily obtained. The result is
 \be
{\cal Z} [\epsilon ] =\sum_{n_1,n_2} \delta_{n_1, n_2}
\frac{\xi_1^{n_1}}{n_1!}\,\,\frac{\xi_2^{n_2}}{n_2!}
= I_0(2\sqrt{\xi_1\xi_2})\sim\frac 1
{ \sqrt{4\pi} (\xi_1\xi_2)^{\frac 1 4}} e^{2\sqrt{\xi_1\xi_2}}
\label{166} \ \ \ \ \ ,
\ee
where $I_0$ is a modified Bessel function, and where again $\sim$
means the limit $L\rightarrow\infty$.
Substituting (A.1) and (A.2) in (A.3) and calculating the expectation value
of the Wilson loop as in section 3  gives  the following result
\be
\lan W (x) \ran_{m,\vartheta=\pi/N} = z^* \lambda^{N/2-1}e^{-\beta F_1}
\label{19}\ \ \ \ \ .
\ee
For $N>2$ the expectation value is zero, while for $N=2$ it is the same as
in the massless case. This is  completely consistent with the results in
the text, since for $N=2$ there are no ''clusters'', but only ''free
charges'', just as in the massless case.  It would be interesting to
find out whether any similar effect would occur in four dimensional
QCD for a finite vacuum angle.

\newpage

\end{document}